\documentclass[
,final            
]
  {aipproc}
\layoutstyle{6x9}
\def\be{\begin{eqnarray} &&}

\def\ee{\end{eqnarray}}
\def\psla{\rlap \slash}
 
\input psfig.sty
\usepackage{graphicx}
\usepackage{epsfig}
\begin{document}
\title{Pion Form Factor in the Light-Front 
\footnote{To appear in the proceedings
''IX Hadrons Physics and VII Relativistic Aspects 
of Nuclear Physics: A joint Meeting on QCD and QGP, 
Hadron-RANP 2004'', 
Angra dos Reis, Rio de Janeiro, Brazil.}}
\author{J.~P.~B.~C.~de~Melo
\footnote{also, LPNHE, Universit\'e P.~\& M.~Curie, 
4 Place Jussieu, F-75252, Paris, France.}
} 
{address={Instituto de F\'\i sica T\'eorica, Universidade  
Estadual Paulista
\\
Rua Pamplona, 145, 
01405-900, S\~ao Paulo, SP., Brazil}
}
\begin{abstract}
The pion electromagnetic form factor is calculated with 
a light-front quark model. 
The "plus" and "minus" components of the 
electromagnetic current are used to calculate 
the  
electromagnetic form factor in the the 
Breit frame with two models for the 
$q\bar{q}$ vertex.  
The light-front constituent quark model describes very well the 
hadronic wave functions for 
pseudo-scalar and vector 
particles. 
Symmetry problems arising in the light-front 
approcah are solved by the pole dislocation method. 
The results 
are compared with new experimental data 
and with other quark models. 
\end{abstract}
\maketitle
\vspace{-1.5cm}
\section{Introduction}
In the light-front approach, the wave 
functions for the hadronic bound states 
are defined in the hypersurface~$x^+=x^0+x^3=0$. 
These wave functions 
are covariant under kinematical light-front 
boosts~(see ref.~\cite{Brodsky98}
for details on light-front field theory). 
Within the light-front constituent quark 
model~({\it LFCQM})~(\cite{Jaus91}~to~\cite{Pacheco2002}), 
the relativistic description of the lightest hadron 
bound state is a natural approach. 
The {\it LFCQM} describes very well the electromagnetic 
proprieties of the
hadronic bound 
states~\cite{Tobias92,Pacheco97,Pacheco99,Ji2001,Pacheco2004,Pacheco2002}, 
but there exist 
some problems related with the breaking of the rotational symmetry 
and there are 
{\it ''zero modes''} for particles of spin-0 and 
spin-1~\cite{Pacheco97,Pacheco99,Ji2001,Pacheco2002}. 
The previous light-front models for the pion~\cite{Pacheco99,Pacheco2002} 
are extended to momentum transfer square 
up to 10 $(GeV/c)^2$ and compared with 
new experimental data, and with the QCD sum rule and 
vector meson dominance models.  
In the references~\cite{Roberts96,Maris2000,Troitsky2001,
Carvalho2004,Desplanques2004}, anothers approaches are considered 
to describe light bound states, like pion and rho mesons. 
\vspace{-.6cm}
\section{Electromagnetic Form Factor}
The ''plus~($+$)'' and ''minus~($-$)''
components of the electromagnetic current, used 
to extract the pion
electromagnetic form factor, are:
\begin{eqnarray}
J^{\pm}_\pi&=& e (p^{+}+p'^{+}) F^{\pm}_\pi(q^2)= 
\nonumber \\ 
& & \hspace{-1.0cm} -2 \imath e
\frac{m^2}{f^2_\pi} 
N_c \int \frac{d^4k}{(2\pi)^4}
\frac{Tr[ \ \gamma^5 S(k-p)\gamma^{\pm}S(k-p')\gamma^5 S(k)\ ]
\Gamma(p,k)\Gamma(p',k)} 
{({k^2-m^2 + \imath \epsilon}) 
((k-p')^2-m^2+ \imath \epsilon) 
((k-p)^2-m^2 + \imath \epsilon)}. 
\label{current} 
\end{eqnarray}
In Eq.~(\ref{current}), 
the Dirac "plus" and ''minus'' matrices are given by   
$\gamma^{\pm}=\gamma^0\pm\gamma^3$~\cite{Brodsky98}, 
the Dirac propagator for fermions is
$ S(p)=1/(\psla{p}-m^2+\imath \epsilon)$, $m$ is 
the constituent quark mass,  
$f_{\pi}$ is the pion decay constant, 
$N_c=3$ is the number of color and the factor 2 
comes from isospin algebra. 
The electromagnetic form factors 
are then calculated in the 
Breit frame with $p^{\mu}=(p_0,-q/2,0,0)$, 
$p^{{\prime}\mu}=(p_0,q/2,0,0)$ 
and with the momentum transfer 
$q^{\mu}=(p^{\prime\mu}-p^{\mu})=(0,q,0,0)$, 
for two models of the 
$\pi-q{\bar q}$ vertex. 
The two vertices considered are 
the nonsymmetric vertex~(NSY) and the 
symmetric one~(SY)~\cite{Pacheco99,Pacheco2002}:
\begin{equation} 
\Gamma^{NSY}(p,k)=
\frac{N}{((p-k)^2-m^2_R+\imath \epsilon)}, 
\label{nosy}
\end{equation}  
\begin{equation} 
\Gamma^{SY}(p,k)=
\frac{N}{(k^2-m^2_R + \imath \epsilon)} + 
\frac{N}{((p-k)^2-m^2_R + \imath \epsilon)}, 
\label{vesy}
\end{equation} 
where $m_R$ is a regulator mass. 
With the nonsymmetric vertex,~(Eq.\ref{nosy}), the 
electromagnetic form factors of the pion, 
for the ''plus'' and ''minus'' components of the 
electromagnetic current, are given by:
{\small
\begin{eqnarray}
F_{\pi}^{+(i)(NSY)}(q^2) & = & 
\frac{m^2}{ p^+ f^2_\pi} N_c 
\int \frac{d^{2} k_{\perp} d x}
{2(2 \pi)^3 x (1-x)^2 } 
\left[-4 \left(\frac{f_1}{x p^+}\right)  
(x p^+ - p^+)^2 + \right. 
\nonumber \\  
&   &  \left.  
(x p^+-2 p^+) 4 f_{1} - x p^+ q^2    \right]        
 \Psi^{*(NSY)}_f(x,k_{\perp})
\Psi^{(NSY)}_i(x,k_{\perp})
\theta(x) \theta(1-x),  
\label{ffplus}
\end{eqnarray} 
\begin{eqnarray}
F_{\pi}^{-(i)(NSY)}(q^2) & = & \frac{m^2}{ p^+ f^2_\pi} N_c 
\int \frac{d^{2} k_{\perp} d x}
{2(2 \pi)^3 x (1-x)^2 } \left[ 
-4 \frac{f_1^2}{x p^+}
-4 p^+ ( 2 f_1+ x p^{+2})\right. 
\nonumber \\ 
&  &  \- \-  \left.   
 + \frac{f_
1}{x p^+}
(4 f_1 + 8 x p^{+2} -q^2)\right]
\Psi^{*(NSY)}_f(x,k_{\perp}) 
\Psi^{(NSY)}_i(x,k_{\perp}) \theta(x)\theta(1-x).
\label{ffminus}
\end{eqnarray}
} 
In Eqs.~(\ref{ffplus})~and~(\ref{ffminus}) 
the light-front coordinates are 
$k^{\mu}=(k^+,k^-,k_\perp)=(k^0+k^3,k^0-k^3,k_{\perp})$
and $f_1=(k_{\perp}^2+m^2)$.
The integration interval (i) is given by 
$0 < k^+ < p^+ $ , i.e., the quark on-shell is 
$k_{on}^{-}=(k_{\perp}^2+m^2)/k^+$, 
$x=k^+/p^+$ and $ 0 \leq x \leq 1 $.
The wave 
function for the pion, $\Psi^{(NSY)}(x,k_{\perp})$, 
is 
\begin{equation}
\Psi^{(NSY)}(x,k_\perp)=\frac{1}{(1-x)}
\frac{N}{(m_\pi^2-M_0^2)(m_\pi^2-{\cal M}^2(m,m_R^2)}, 
\end{equation} 
here ${\cal M}^2(m^2,m_R^2)=
(k_{\perp}^2+m^2)/x+ ((p-k)_{\perp}^2+m_R^2)/(1-x)-p^2_{\perp}$ 
and $M_0^2={\cal M}^2(m^2,m^2)$.
The charge conditions,~$F^{\pm}_{\pi}(0)=1$, 
determine the normalization constant $N$. 
However, for the minus component, another contribution exists, 
i.e, the non-valence or the 
pair term contribution~\cite{Pacheco97,Pacheco99} in the 
second interval (ii), $p^+<k^+<p^{{\prime}+}$, where 
$p^{{\prime} + } = p^{+}+\delta$, 
~(see for details the {\it ''pole dislocation method''} 
in \cite{Pacheco99,Pacheco98,Naus98,Pacheco992}). 
This second contribution to the electromagnetic form factor, 
obtained in the limit $\delta \rightarrow 0$, is:
\begin{equation}
F^{-(ii)(NSY)}_{\pi}(q^2)=
\frac{m^2 N_c N^2}{p^+ f^2_{\pi}}
\int \frac{d^{2} k_{\perp}} {2(2 \pi)^3  } 
\left[  \frac{m^2_{\pi} - \frac{q^2}{2}}{p^+} \right ]  
\sum_{i=2}^5 \frac{ln(f_i)}
{ \prod_{i=2, \ i \neq j}^{5} 
(-f_i+f_j)},  
\end{equation}
where $f_2=(p-k)_{\perp}^2+m^2$, 
$f_3=(p^{'}-k)_{\perp}^2+m^2$, $f_4=(p-k)_{\perp}^2+m^2_R$
and $f_5=(p^{\prime}-k)_{\perp}^2+m^2_R$. 
We then recover the Lorentz covariance of 
the electromagnetic current:
\begin{equation}
F^{+(i)(NSY)}_{\pi}(q^2)=F^{-(i)(NSY)}_{\pi}(q^2)+
F^{-(ii)(NSY)}_{\pi}(q^2).
\end{equation} 
For the ''plus'' component of the electromagnetic 
current, 
the pair terms go to zero and then do not give any 
contribution to the electromagnetic 
form factor~\cite{Pacheco99}.

With the symmetric vertex,~Eq.(\ref{vesy}), 
the valence wave function of the pion is:
\small{
\begin{equation}
\Psi^{(SY)}(x,k_\perp)=
\left[  \frac{  N }
{(1-x)(m^2_{\pi}-{\cal M}^2(m^2, m_R^2))} 
+\frac{ N }
{x(m^2_{\pi}-{\cal M}^2(m^2_R, m^2))}  \right]
\frac{1}{m^2_\pi-M^2_{0}},
\end{equation}
}
and the 
electromagnetic form factor of the pion is:
\small{
\begin{equation}
F_\pi^{(SY)}(q^2)= 
\frac{m^2 N_c}{p^+f_{\pi}^2}
\int \frac{ d^{2} k_{\perp}}{2 (2\pi)^3 }
\int_0^{1} \- \- dx
\nonumber \\
\left[ k^-_{on}p^{+ 2} +  
\frac14 x p^{+} q^2 
\right ]
 \frac{
\Psi^{*(SY)}_{f}(x,k_\perp)
\Psi^{(SY)}_{i}(x,k_\perp)}
{x (1-x)^2}. 
\end{equation}}
The normalization
constant ${ N }$ is determined from the charge condition
$F^{+}_\pi(0)=1$. In the previous work~\cite{Pacheco2002}, 
a general frame was considered, but in the present work, 
only the Breit 
frame is used, with the condition $q^+=0$, 
in order to compare with 
the nonsymmetric vertex case~\cite{Pacheco99}. 
A more complete work, studying the differences 
between the nonsymmetric and symmetric vertices, is in 
preparation~\cite{Pacheco20042}.
\vspace{-0.5cm}
\section{results and conclusion}
For the nonsymmetric vertex model, 
$m_R$ is chosen to be 0.946 GeV in order to 
fit the pion experimental radius
$r_{exp}=0.67\pm0.02$~fm~\cite{Amendolia84}. In the case of 
the symmetric vertex model, $m_R$ is fixed to 
0.6 GeV in order to 
reproduce the value of $f_{\pi}^{exp}$=92.4~MeV. 
We take $m_{\pi}=$0.140~MeV and the same quark mass
$m$=0.220 GeV for both vertex model calculations. 
In Fig.~1 it can be seen that the pion 
electromagnetic form factor, for both vertex models 
in the light-front approach, compares very well with the 
experimental data~\cite{Amendolia84,Volmer2001,Frascati2001}
and with the QCD sum rule and vector meson dominance models. 
In r\'esum\'e, 
the light-front models presented here, work very well at low 
and high energies, however, 
differences, under investigation~\cite{Pacheco20042}, 
exist between the two vertex models. 
For both vertices considered~\cite{Pacheco99,Pacheco2002}, 
it is important 
to take correctly the valence and 
non-valence (pair terms) contributions to the 
matrix elements of the electromagnetic current. 
\begin{figure}
\includegraphics[height=.5\textheight]{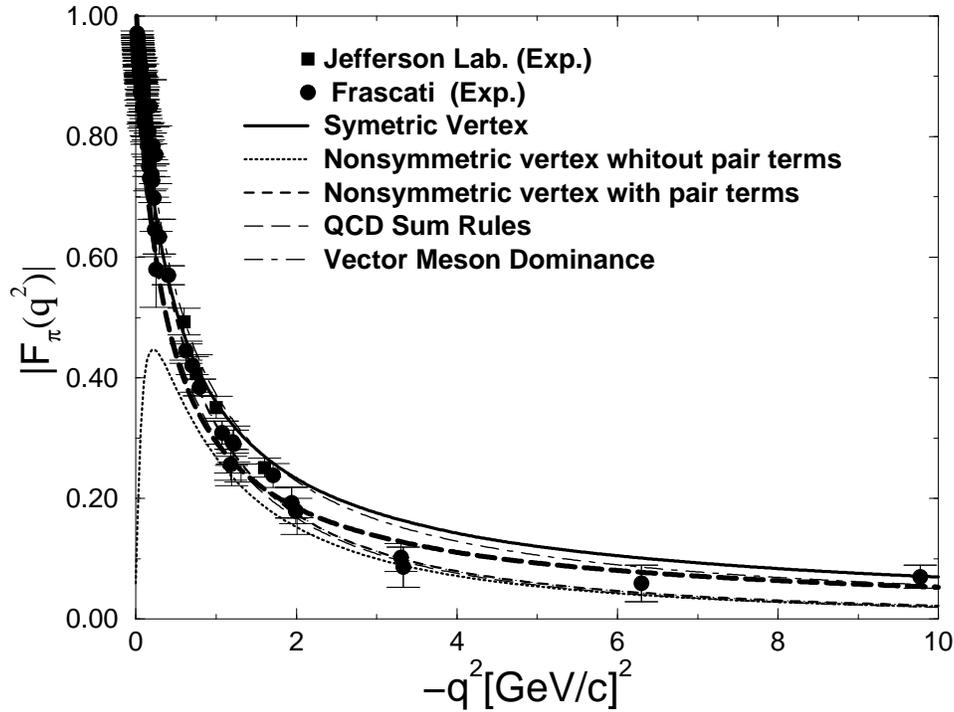}
  \caption{Pion electromagnetic form factor, calculated in the 
light-front approach and compared with 
QCD sum rules~\cite{Nesterenko82}, 
vector meson dominance(VMD)~\cite{Conne97} and 
experimental data~\cite{Amendolia84,Volmer2001,Frascati2001}. }
\end{figure}
\vspace{-1.008cm}
\begin{theacknowledgments}
This work was supported by the Brazilian 
Agency Funda\c{c}\~ao de Amparo \'a Pesquisa do 
Estado de S\~ao Paulo, FAPESP. The author thanks 
T. Frederico, E. Pace and G. Salm\'e 
for fruitful discussions and collaboration. 
The author thanks also B. Loiseau for 
critically reading the manuscript. 
\end{theacknowledgments}
\bibliographystyle{aipproc}   
\bibliographystyle{aipprocl} 
\IfFileExists{\jobname.bbl}{}
  {\typeout{}
   \typeout{******************************************}
   \typeout{** Please run "bibtex \jobname" to optain}
  \typeout{** the bibliography and then re-run LaTeX}
  \typeout{** twice to fix the references!}
   \typeout{******************************************}
   \typeout{}
 }

\end{document}